\documentclass{ifacconf}

\usepackage{bm}
\usepackage{float}
\usepackage{algorithmic}
\usepackage{bm}
\usepackage{amsmath} 
\usepackage{amssymb}  
\usepackage{graphicx}      
\usepackage{natbib}        
\begin{document}
\begin{frontmatter}

\title{Smart Meters Integration in Distribution System State Estimation with Collaborative Filtering and Deep Gaussian Process\thanksref{footnoteinfo}} 

\thanks[footnoteinfo]{This work was supported by the National Key R\&D Program of China(2020YFB0906000, 2020YFB0906005). Corresponding author: Ye Guo (guo-ye@sz.tsinghua.edu.cn).}

\author[First]{Yifei Xu, Ye Guo, Wenjun Tang} 
\author[Second]{Hongbin Sun} 
\author[Third]{Shiming Li, Yue Dai}

\address[First]{Tsinghua-Berkeley Shenzhen Institute, Tsinghua University, Shenzhen, China (e-mail: guo-ye@sz.tsinghua.edu.cn).}
\address[Second]{Department of Electrical Engineering, Tsinghua University,
Beijing, China (e-mail: shb@tsinghua.edu.cn)}
\address[Third]{Electrical Dispatch and Control Center, Guangdong Power Grid,
Guangzhou, China (e-mail: lishiming@gddd.csg.cn, 13450243047@163.com)}

\begin{abstract}                
The problem of state estimations for electric distribution system is considered. A collaborative filtering approach is proposed in this paper to integrate the slow time-scale smart meter measurements in the distribution system state estimation, in which the deep Gaussian process is incorporated to infer the fast time-scale pseudo measurements and avoid anomalies. Numerical tests have demonstrated the higher estimation accuracy of the proposed method.
\end{abstract}


\end{frontmatter}

\section{Introduction}
The electric distribution system state estimation (DSSE) plays a paramount role in the distribution management system (DMS). It achieves the real-time monitoring of system states so that system operators can make their dispatch decisions efficiently. Typically, a state estimation program in the electric transmission system utilizes redundant measurements to filter the most likely system states. However, in the distribution system, the measurements from remote terminal units (RTU) and micro phasor measurement units($\mu$PMU) are limited, which cannot guarantee the observability [\cite{abur2004power}] of the distribution system. A conventional solution is to incorporate the pseudo measurements to make the system observable [\cite{primadianto2016review}]. Unfortunately, the pseudo measurements are usually of poor accuracy. Several forecasting techniques have been studied to improve the quality of the pseudo measurements [\cite{wu2012robust,hayes2014closed}].

With the advent of the smart grid, smart meters are deployed vastly to report the demand of customers, which makes the monitorization of unobservable areas possible. However, there are several prominent problems that hinder its application in the DSSE. First, the data of the smart meters are updated less frequently compared to that of RTUs and $\mu$PMUs. The sampling interval of RTUs or $\mu$PMUs is in seconds or milliseconds level, while the sampling interval of smart meters is about 15-60 minutes. Second, smart meters record the cumulative energy consumption or the average power, while RTUs and $\mu$PMUs record the real-time power and voltage.

Several methods have been proposed to exploit the smart meter data in the DSSE. For example, the credibility of the non-synchronized smart meter data is modeled by the variance of the measurements in \cite{alimardani2015distribution}. A two time-scale state estimation method is proposed in \cite{gomez2014state}. Extrapolation and interpolation are incorporated to obtain the fast time-scale pseudo measurements. In \cite{mestav2019bayesian}, the smart meter measurement is modeled as a stationary process with a Gaussian mixture distribution, and the probability distribution is obtained through the $AR$-$k$ process. However, most of the studies did not consider the cumulative energy measurements and load information from other nodes, which may not reflect the real-time characteristics of the smart meter node.

Collaborative filtering (CF) can be used to estimate the fast time-scale measurements of the smart meter nodes by considering load correlations. CF is a technique widely used in recommendation systems. It makes automatic predictions (filtering) about the interests of a user by collecting preferences or taste information from many users(collaborating) [\cite{su2009survey}]. In \cite{8412100}, CF has been used to construct the electricity market recommendation system in power systems.

To reflect the load features and improve the performance of CF, the load forecasting technique can be further incorporated. The Gaussian process (GP) has demonstrated its effectiveness in modeling time series. Given that power system state estimation measurements are inherently time series, it is reasonable to model them as a GP. Consequently, Gaussian process regression (GPR) can be applied to predict the time series. GPR is a non-parametric machine learning method based on Bayesian theory. It can learn the underlying distribution from limited training data, which helps evaluate the uncertainties and promote the system's operation. However, the performance of the GPR is limited by the selection of the kernel function, which hinges on the specific characteristic of the dataset. The deep Gaussian process(DGP) is a hierarchical paradigm of the GP that can handle this problem by stacked layers which are similar to deep neural networks [\cite{salimbeni2017doubly}]. The GPR has already been applied in power system load forecasting [\cite{cao2021robust}], optimal power flow [\cite{pareek2020gaussian}] and voltage control [\cite{ye2021global}].

In this paper, a collaborative filtering approach is proposed to integrate smart meter measurements in the DSSE. Based on the correlation analysis of the measurements, the CF is used to infer the fast time-scale pseudo measurements. The measurements are modeled as the Gaussian process. The DGP is further introduced to extract the features of the measurements and make the algorithm more robust. With the information from both smart meters and real-time measurements, the proposed method can achieve better estimation accuracy.

\section{The model of smart meters}
We first review measurement conditions in distribution systems. In general, there are two types of measurements: One is real-time measurements from remote terminal units and micro phasor measurement units. The other is the advanced measurement infrastructure (AMI) with smart meter measurements, which are updated every 15-60 minutes. The following parts of this section will elaborate the technique to integrate the smart meter data in the DSSE.

\subsection{Smart Meter Measurement Model}
\begin{figure*}[ht]
   \centering
   \includegraphics[width=1\textwidth]{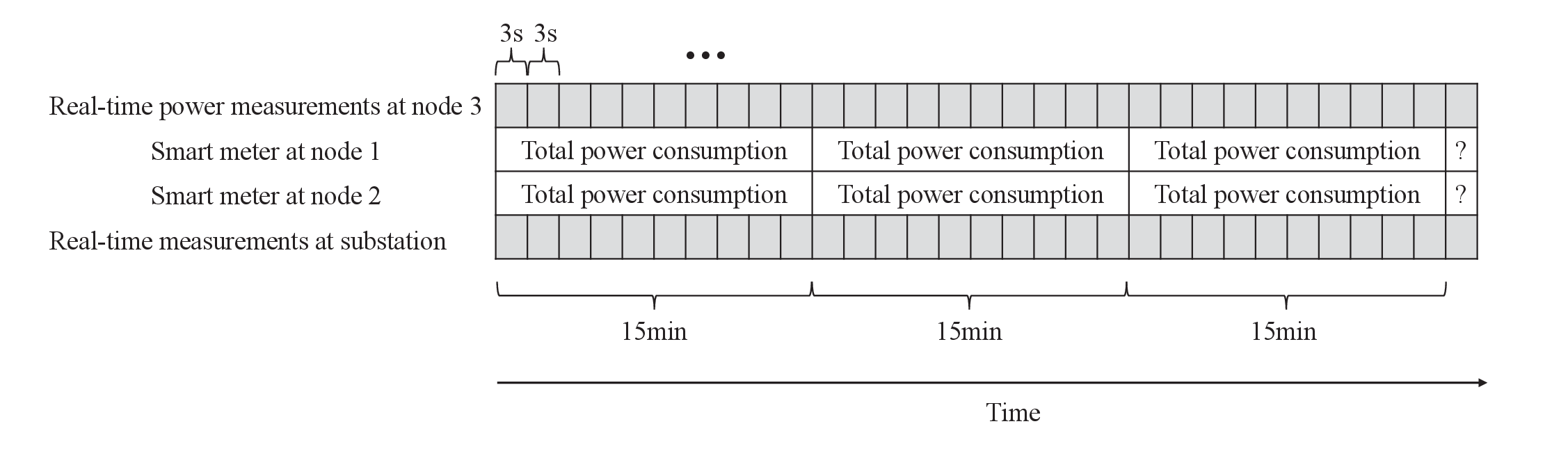}
   \caption{\small Different time scales for smart meter measurements and real-time  measurements}\label{fig:1}
 \end{figure*}
 
 Typically, a smart meter measures the cumulative energy consumption (kWh) every 15-60 minutes, as is depicted in Fig.\ref{fig:1}. For convenience, we assume a series of discrete time intervals. Real-time measurements are updated once per interval and smart meter measurements are updated once every $T$ intervals. Then for every smart meter measurement, we have
 \begin{equation}
 W_{i}=\sum_{t=1}^TP_{i,t}\Delta t,
 \label{sp}
 \end{equation}
 where $W_{i}$ is the smart meter measurement of node $i$, and $P_{i,t}$ is the real-time measurement of node $i$ at time $t$. $\Delta t$ is the length of each time interval.
 
 \textit{\textbf{Assumption:}  
 $(P_{i,t})$ is a Gaussian process, which makes $W_i$ also a Gaussian process.}
 
 A time continuous stochastic process $P_{i,t}$ is Gaussian if and only if for every finite set of $t_1,...,t_k$, $(P_{i,t_1},...,P_{i,t_k})$ is a multivariate Gaussian random variable.
 
 For multiple random variables satisfying the Gaussian distribution, their sum still follows the Gaussian distribution. Thus, $W_i$ is also a Gaussian process.
 
 In Section III, we will utilize the Gaussian process regression to analyze the intrinsic features and predict the fast time-scale pseudo measurements from the smart meters.
 
 \subsection{Collaborative Filtering}
 According to our experiences in the recommendation system, if person A has the same opinion as person B on an issue, A is more likely to have B’s opinion on a different issue than that of a randomly chosen person. Power consumptions, in the end, are personal behaviors. Therefore, if node 1 with the smart meter has similar cumulative load patterns with node 2, then the fast time-scale characteristic of node 1 will also resemble that of node 2. Although the fast time-scale pattern of the bus with smart meters only is unknown, we can presume that it is likely to have similar patterns with other kindred buses equipped with RTUs/$\mu$PMUs.
 
 Since there exist two different time scales, we have to do some transformations first to compare their similarities with a uniform standard. Here we convert the fast-scale measurements to the cumulative measurements. According to (\ref{sp}), the cumulative power consumption for the $T$ intervals can be calculated. Then the correlations of different loads can be analyzed through the cumulative energy consumption.
 \begin{figure}[ht]
   \centering
   \includegraphics[width=0.37\textwidth]{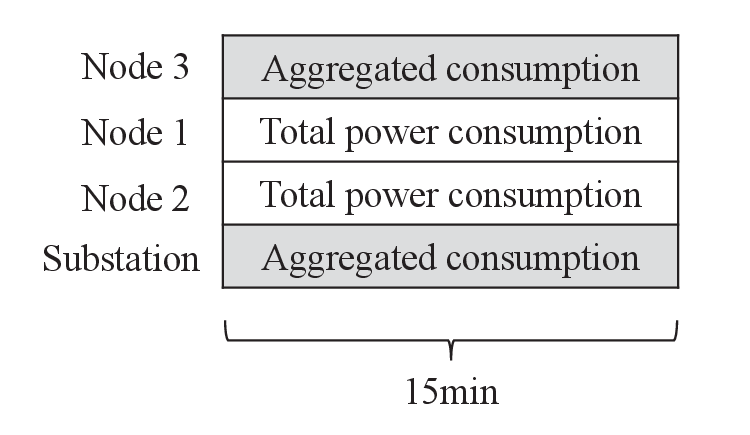}
   \vspace{-1em}
   \caption{\small Load aggregation for different nodes}\label{fig:2}
 \end{figure}
 
 We introduce the Pearson correlation to evaluate the similarity between different buses, which is given as follows:
 \begin{equation}
 r_{i,j} = \frac{\sum_{t\in l_{i,j}}(w_{i,t}-\bar{w_i})(w_{j,t}-\bar{w_j})} {\sqrt{\sum_{t\in l_{i,j}}(w_{i,t}-\bar{w_i})^2}\sqrt{\sum_{t\in l_{i,j}}(w_{j,t}-\bar{w_j})^2}}
 \label{pearson}
 \end{equation}
 where $l_{i,j}$ denotes the time-instant index of the data. $w_{i,t}$ and $w_{j,t}$ are the cumulative energy consumption of the $ith$ and $jth$ node at time $t$, and $\bar{w_i}$ and $\bar{w_j}$ denote the average consumption of all time, regarding the $ith$ and $jth$ node. The value range of (\ref{pearson}) is between -1 and 1. A value close to 1 means that node $i$ and $j$ are strongly positive-correlated. 
 
 Based on the theory of collaborative filtering, we first propose the following estimate of fast-scale measurements:
 \begin{equation}
 p_{i, t}=\bar{p}_{i}+\frac{\sum_{j \neq i}\left(p_{j, t}-\bar{p}_{j}\right) r_{i, j}}{\sum_{j \neq i}\left|r_{i, j}\right|}
 \label{eq:CF}
 \end{equation}
 where $p_{i,t}$ is the fast-scale pseudo estimate of the smart meter node $i$ at time $t$. $\bar{p}_i$ and $\bar{p}_j$ are the average power measurements of all time, regarding node $i$ and $j$. $p_{j,t}$ is the real-time measurement of node $j$ at time $t$.
 
 The basic CF method in (\ref{eq:CF}) uses the real-time measurement $p_{j,t}$ to estimate $p_{i,t}$. However, there are some drawbacks in this implementation. First, the node with smart meters may not be observable with real-time measurements. Although it is possible to use pseudo measurements to recover the observability, this may still create critical measurements [\cite{abur2004power}] whose gross errors are impossible to identify. The gross error in $p_{j,t}$ will cause a bias on $p_{i,t}$. Second, the size of every window is ambiguous, and a simple average cannot reflect the dynamic evolution of power systems.
 
 In order to tackle these obstacles, the deep Gaussian process regression (DGPR) is introduced to avoid the anomaly in $p_{j,t}$ and extract load features. The DGPR makes the algorithm more robust since it uses the Gaussian mean rather than the raw measurements, which is similar to the idea of forecasting-aided state estimation and innovation analysis [\cite{do2009forecasting}]. We care about the long-time trends and similarities of other nodes rather than a snapshot. The mean value can better reflect the trend and avoid anomalies.
 
 Therefore, we further propose a CF method with deep Gaussian process (CF-DGP), which is given as follows.
 \begin{equation}
 p_{i,t} = \frac{\tilde{W}_{i,t_S+T}}{T}+\frac{\sum_{j\neq i}r_{i,j}(\tilde{p}_{j,t}-\tilde{p}_{j,t_S+T})}{\sum_{j\neq i}\mid r_{i,j}\mid }
 \label{eq:CF-DGP}
 \end{equation}
 where $p_{i,t}$ is the fast-scale pseudo measurement of the smart meter node $i$ at time $t\subseteq[t_S, t_S+T)$. $t_S$ is the sampling time of the latest available smart meter measurements. $\tilde{W}_{i,t_S+T}$ is the Gaussian predictive mean of the Gaussian process $W_i$ at time $t_S+T$. $\tilde{p}_{j,t}$ and $\tilde{p}_{j,t_S+T}$ are the Gaussian means of $(P_{j,t})$ at time $t$ and $t_S+T$, respectively. 
 
 The first term on the right-hand side in (\ref{eq:CF}) and (\ref{eq:CF-DGP}) is an average term that reflects the historical load level of node $i$. The second term is a trend term that estimates the trend information from other correlated nodes. 
 
\begin{figure}[ht]
   \centering
   \includegraphics[width=0.47\textwidth]{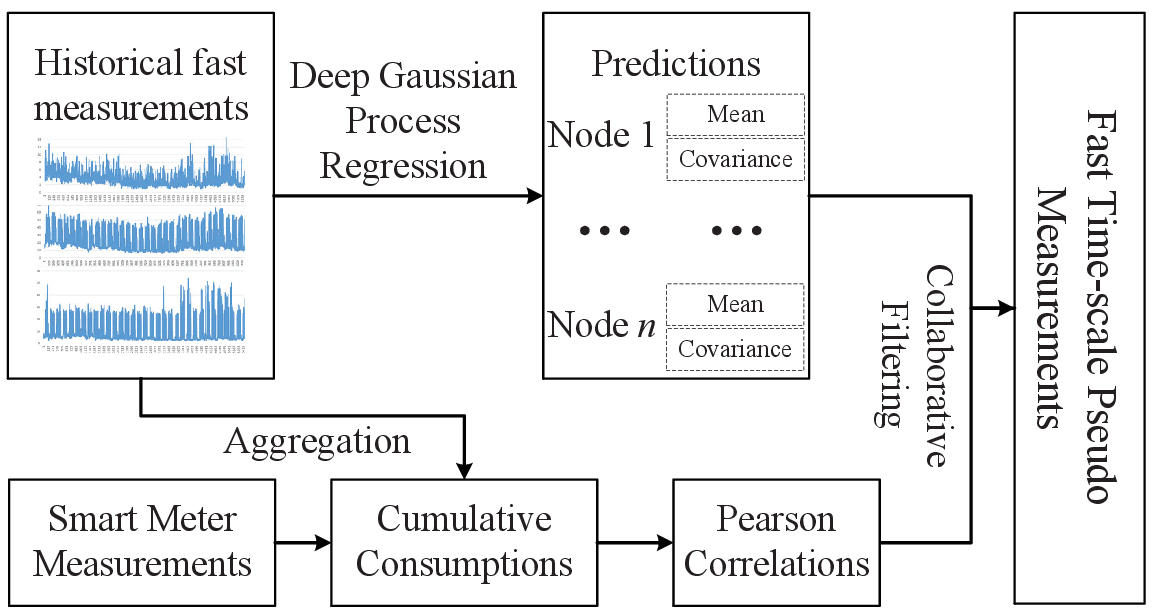}
   \vspace{-1em}
   \caption{\small Schematic of the proposed CF-DGP}\label{fig:3}
 \end{figure}

 The structure of the proposed CF-DGP method is summarized in Fig.\ref{fig:3}.

  \section{The DGP model for CF-DGP}
  In this section, we will elaborate the deep Gaussian process regression [\cite{damianou2013deep}], a non-parametric machine learning model mentioned in Fig.\ref{fig:3}. We use DGP regression to forecast the Gaussian mean and covariance of a Gaussian process.
  
  Given a set of $N$ observations $\bm{Y}=(y_1,...,y_N)^T$ with the inputs $\bm{X}= (\bm{x_1},...,\bm{x_N})^T$, we place a zero mean Gaussian process prior on the latent function $f(\bm{x})$ with a covariance function $k(\bm{x_i},\bm{x_j})$:
  \begin{equation}
  f(\bm{X})\sim GP\left(\bm{0},k(\bm{x_i},\bm{x_j})\right)
  \label{2.1}
  \end{equation}
  Here we assume that the noise in the observations is additive, independent and Gaussian. Then the relationship between the latent function values and the observations can be modeled as:
  \begin{equation}
  \bm{Y}=f(\bm{X})+\bm{\varepsilon} \quad \text{with}\quad \bm{\varepsilon} \sim N(\bm{0},\sigma_n^2\bm{I})
  \end{equation}
  where $\bm{\varepsilon}$ is the Gaussian noise; $\sigma_n^2$ is the variance; $\bm{I}$ is the identity matrix.
  
  Based on the Bayesian theory, the Gaussian prior distribution of $\bm{Y}$ is given by
  \begin{equation}
  p(\bm{Y})=N(\bm{0},k(\bm{X,X})+\sigma_n^2\bm{I})
  \end{equation}
  For $\bm{x^*}$, which is the point to be estimated, and the associated output $\bm{f^*}$, the joint distribution of the training output $\bm{Y}$ and the test output $\bm{f^*}$ are
  \begin{equation}
  \begin{bmatrix}
  \bm{Y}\\
  \bm{f^*}
  \end{bmatrix}\sim
  N\left(\bm{0},\begin{bmatrix}
  K_{\bm{XX}}+\sigma_n^2\bm{I} & K_{\bm{Xx^*}}\\
  K_{\bm{x^*X}} & K_{\bm{x^*x^*}}
  \end{bmatrix}\right)
  \label{eq:joint}
  \end{equation}
  where $K_{\bm{XX}}$ denotes the covariance matrix on the input variables, and the entries of the covariance matrix is given by the covariance function. The posterior distribution of $\bm{f^*}$ is consequentially given by the conditioning of (\ref{eq:joint}).
  \begin{equation}
  p(\bm{f^*}|\bm{x^*,X,Y})\sim N(\bm{\hat{m}},\bm{\hat{K}})
  \label{eq:pred}
  \end{equation}
  \vspace{-1em}
  \begin{equation}
  \begin{gathered}
  \bm{\hat{m}}=K_{\bm{x^*x}}^T(K_{\bm{XX}}+\sigma_n^2\bm{I})^{-1}K_{\bm{x^*x^*}}\\
  \bm{\hat{K}}\!=\!K_{\bm{x^*x^*}}\!-\!K_{\bm{x^*X}}(\!K_{\bm{XX}}\!+\!\sigma_n^2\bm{I})^{\!-\!1}K_{\bm{Xx^*}}
  \end{gathered}
  \end{equation}
  
  The selection of the covariance function is empirical, which hinges on the specific features of the dataset. The most widely used function is the Radial Basis Function (RBF), which is defined as follows:
  \begin{equation}
    k(\bm{x_i,x_j})=\sigma_{f}^{2} \exp \left(-\frac{1}{2}(\bm{x_i}-\bm{x_j})^{T} M^{-1}(\bm{x_i}-\bm{x_j})\right)
  \end{equation}
  where $\sigma_f^2$ and $M$ are learnable variance parameters. We can infer the predictive distribution at $\bm{x^*}$ with the mean and covariance function. The variance allows us to quantify the forecasting uncertainties.
  
  The main obstruction for applying GPR in a large training set is its computation complexity, which is $O(N^{3})$ and $N$ is the number of training inputs. To cope with this limitation, the sparse Gaussian process(SGP) [\cite{snelson2006sparse}], a variant of the GPR, is adopted in this work and the complexity can be reduced to $O\!(N\!M^2\!)$, where $M$ is the number of pseudo inputs and $M\!\ll \!N$.
  
  In SGP, the training set $\bm{X}$ is augmented by the pseudo input $\bm{Z}$, with function values $\bm{u}\!=\!f(\bm{Z})$. $\bm{u}$ is subsequently marginalized to obtain the posterior distribution. However, the marginalization of the non-linear covariance function is intractable. Therefore, variational inference is introduced to approximate the variational posterior $q(\bm{u,f})=p(\bm{f}|\bm{u})p(\bm{u})$ through the minimization of Kullback-Leibler(K-L) divergence between the true posterior $p$ and the variational posterior $q$, where $p(\bm{f}|\bm{u})$ is given by the Gaussian process and $q(\bm{u})=N(\bm{m},\bm{S})$. Here the conditioning on the $\bm{X,Y,Z}$ is omitted to simplify the expression. The closed-form marginal distribution can be expressed as
  \begin{equation}
      q(\bm{f})=\int p(\bm{f}|\bm{u}) q(\bm{u}) d \bm{u}=N(\bm{\hat{m}},\bm{\hat{S}})
  \end{equation}
  And the mean and covariance functions are given by
  \begin{equation}
  \begin{gathered}
  \bm{\hat{m}}=K_{\bm{XZ}} K_{\bm{ZZ}}^{-1} \bm{m}\\ \hat{\bm{S}}=K_{\bm{X} \bm{X}}-K_{\bm{X} \bm{Z}} K_{\bm{Z} \bm{Z}}^{-1}\left(K_{\bm{ZZ}}-\bm{S}\right) K_{\bm{ZZ}}^{-1} K_{\bm{XZ}}^{T}
  \end{gathered}
  \label{eq:SGP}
  \end{equation}
  
  According to (\ref{eq:pred}), the performance of the Gaussian process regression is closely related to the choice of the covariance function. To eliminate the ambiguity brought by different covariance functions, the deep Gaussian process can be formulated by stacking SGPs:
  \begin{equation}
  \bm{y}=f_{L}\left(f_{L-1}\left(\cdots f_{1}(\bm{X})\right)\right)+\bm{\varepsilon}
  \label{eq:DGP1}
  \end{equation}
  where $f_l(\cdot)$ is the $l$th layer's latent function.
  
  The output of the $l$th layer is used as the input of the $(l+1)$th layer. The layers between the input and output layers are known as the hidden layers. The output of the hidden layer and the pseudo input are marginalized to obtain the likelihood. Double stochastic variational inference [\cite{salimbeni2017doubly}] is used to achieve accurate approximations while maintaining the conditional structure. The variational parameters are optimized by maximizing the evidence lower bound(ELBO) of the marginal likelihood $p(\bm{y})$:
  \begin{equation}
    \mathcal{L}=\mathbf{E}_{q\left(\bm{f}_{L}\right)}\left[\log p\left(\bm{y} \mid \bm{f}_{L}\right)\right]-\sum_{l=1}^{L} K L\left[q\left(\bm{u}_{l}\right) \| p\left(\bm{u}_{l}\right)\right]
    \end{equation}
  where $\bm{f}_L$ is the function value $f_L\left(\bm{f_{L-1}}\right)$ of the $L$th layer, and $\bm{u_l}$ is the function value $f_l\left(\bm{Z_l}\right)$, with the pseudo input $\bm{Z}_l$ of the $l$th layer. $KL(\cdot)$ denotes the K-L divergence.
  
  For the test input $\bm{x^*}$, the prediction is computed by first sampling a random variable $\varepsilon_{l} \sim N(0,1)$ for each layer, and then computing the layer output as 
  \begin{equation}
  \bm{\widehat{f}_{l}}=\widehat{\bm{\mu}}_{l}\left(\bm{\widehat{f}_{l-1}}\right)+\varepsilon_{l} \sqrt{\widehat{K}_{l}\left(\bm{\widehat{f}_{l-1}, \widehat{f}_{l-1}}\right)}
  \end{equation}
  where the $\widehat{\bm{\mu}}_{l}(\cdot)$ and $\widehat{K}_{l}(\cdot, \cdot)$ are given in (\ref{eq:SGP}). By stacking each layer, finally we can obtain the predictive value in (\ref{eq:DGP1}).
  
  $\textit{Remark}:$ The Gaussian mixture model is used to model the electricity load in some papers [\cite{singh2009statistical, li2019enhanced}], while we adopt the deep Gaussian process in this work. Note that the Gaussian mixture model focuses on modeling the distribution of a single stochastic variable while the Gaussian process regression tries to infer the multivariate distribution of a stochastic process. Every measurement is regarded as a stochastic variable in the GPR.

  \section{Numerical studies}
\begin{figure}[ht]
    \centering
    \includegraphics[width=0.4\textwidth]{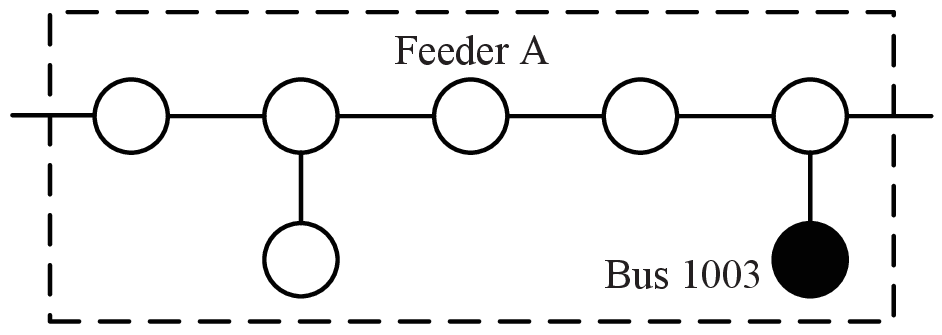}
    \caption{\small The topology of the testbed}\label{fig:topo}
  \end{figure}

  In this section, we test the proposed method on a testbed for the distribution systems analysis [\cite{bu2019time}] to illustrate the effectiveness. This test system is a real distribution grid located in the Midwest U.S. It belongs to a municipal utility and is a fully observable network with smart meters installed at all customers. The data ranges from January 2017 to December 2017. The available smart meters data contains hourly energy consumption(kWh) of 1120 customers. The slow and fast time scales are set as $T_S=1h$ and $T_F=5min$ in this paper, respectively. In the following test, the fast time-scale measurements associated with Bus 1003 are substituted by the cumulative energy measurements. The topology of the selected subsystem is shown in Fig.\ref{fig:topo}.
  
  First, we do the DGP regression for smart meter measurements to extract their intrinsic features. The stochastic variational Gaussian process(SVGP) [\cite{salimbeni2017doubly}] is used for comparison. The code is written in Python with GPyTorch. The model is trained on Nvidia RTX3060 GPU with 12 gigabytes of memory. The data from June 1st to June 18th is used to train and validate the model, and the data from June 19th to June 30th is used to test the performance. The comparison metric is defined as follows.
  \begin{equation}
  \mathrm{MAPE}\!=\!\frac{100\%}{N} \sum_{t=1}^{N}\left|\frac{\tilde{p}_{i,t}\!-\!p_{i,t}}{p_{i,t}}\right|
  \label{eq:MAPE}
  \end{equation}
  where $N$ is the number of sampling points. $\tilde{p}_{i,t}$ and $p_{i,t}$ are the predictive means and the actual measurements of node $i$ at time $t$.
  \begin{figure}[ht]
    \centering
    \includegraphics[width=0.5\textwidth]{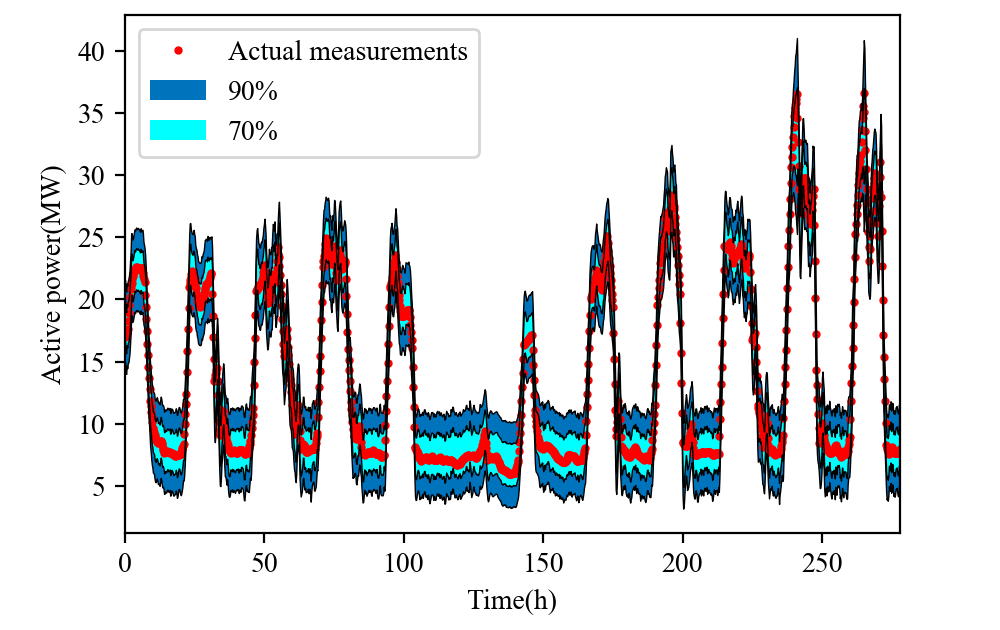}
    \vspace{-2em}
    \caption{\small The DGP regression of Bus 1004}\label{fig:DGP}
  \end{figure}
  
  \begin{figure}[ht]
    \centering
    \includegraphics[width=0.5\textwidth]{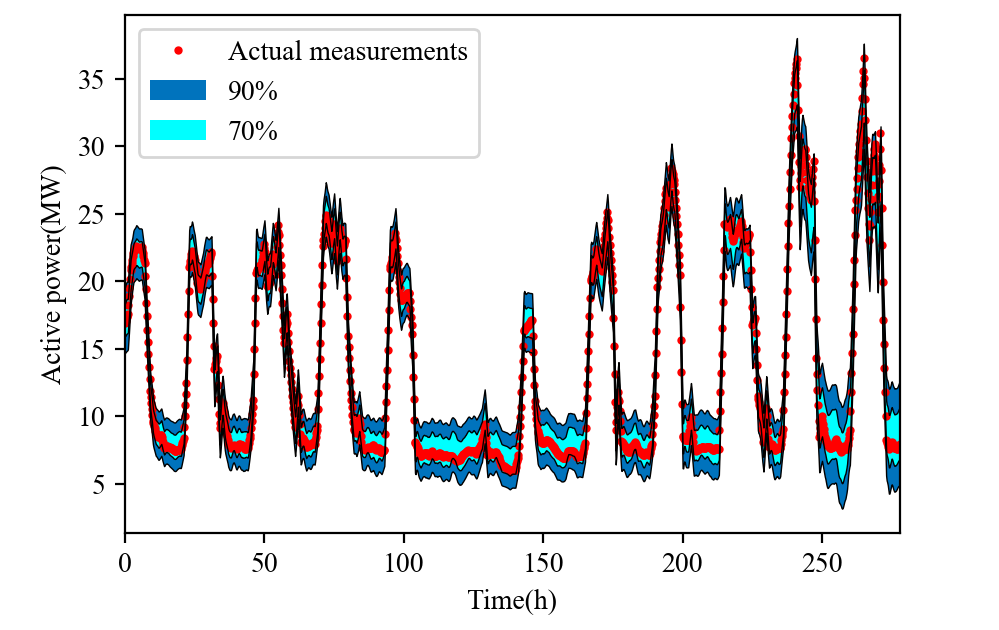}
    \vspace{-2em}
    \caption{\small The SVGP regression of Bus 1004}\label{fig:SVGP}
  \end{figure}
  
  \begin{table}[ht]
  \renewcommand{\arraystretch}{1.2}
  \centering
  \caption{MAPE achieved by SVGP and DGP on validation/test data}
  \setlength{\tabcolsep}{8mm}{
  \begin{tabular}{ll}
  \hline
  Approaches&MAPE(validation/test)\\
  \hline
  SVGP& 3.61\%/5.12\%\\
  DGP&3.15\%/4.08\%\\
  \hline
  \end{tabular}\label{table:gpcompa}}
  \end{table}
  
  The regression results in Figure \ref{fig:DGP} and \ref{fig:SVGP} demonstrate that modeling smart meter measurements as a Gaussian process and performing DGPR is reasonable given limited training data and increased uncertainty. DPG outperforms SVGP in estimation accuracy due to the hierarchical structure, while SVGP may offer a smaller confidence interval.
  
\begin{figure}[ht]
    \centering
    \includegraphics[width=0.43\textwidth]{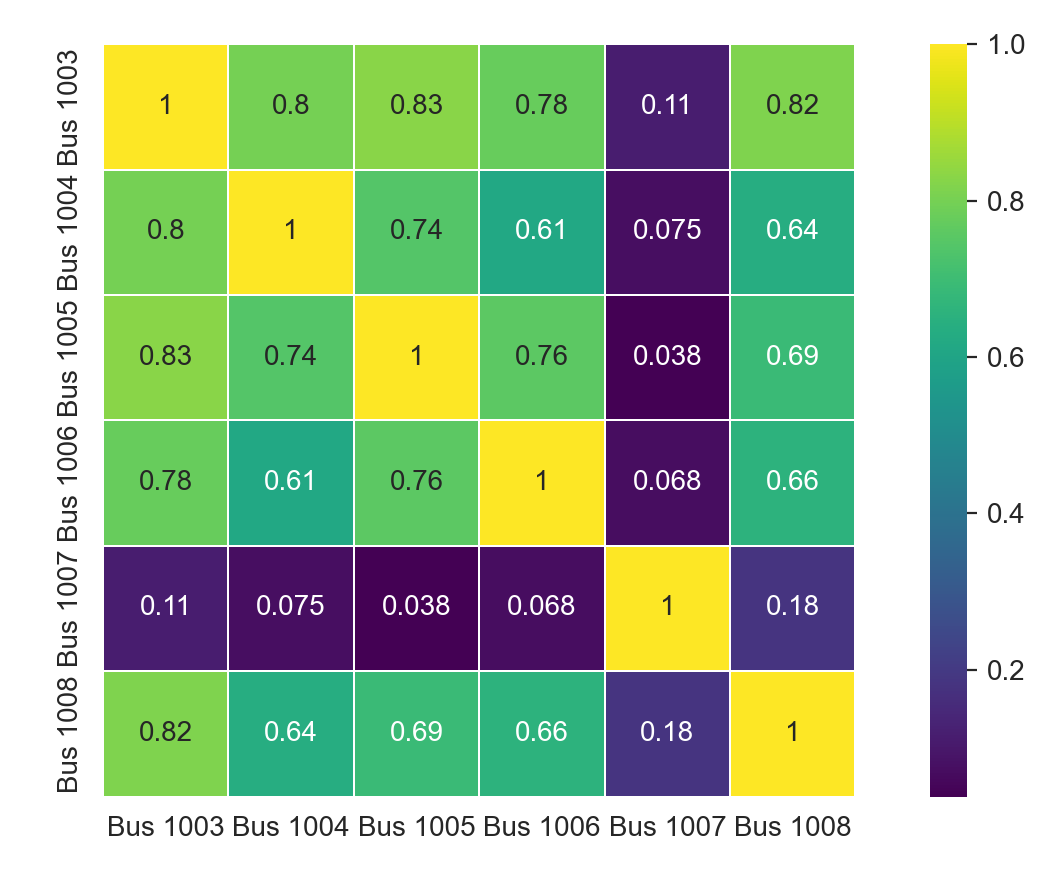}
    \vspace{-1em}
    \caption{\small The Pearson correlation of different buses}\label{fig:pearson}
  \end{figure}

  According to (\ref{pearson}), we analyze the Pearson coefficients of different buses to perform the collaborative filtering. The Pearson correlation of Feeder A is shown in Fig.\ref{fig:pearson}.
  
  From Fig.\ref{fig:pearson}, we can conclude that there is a correlation between different nodes, and it is reasonable to analyze the inherent load patterns of different nodes using CF techniques. However, for some buses, their load patterns are not kindred to other buses, such as Bus 1007. If their fast time-scale measurement is unknown, the proposed method may not have good performance since it is hard to implement the filtering without other information.
  
  \begin{figure}[ht]
    \centering
    \includegraphics[width=0.5\textwidth]{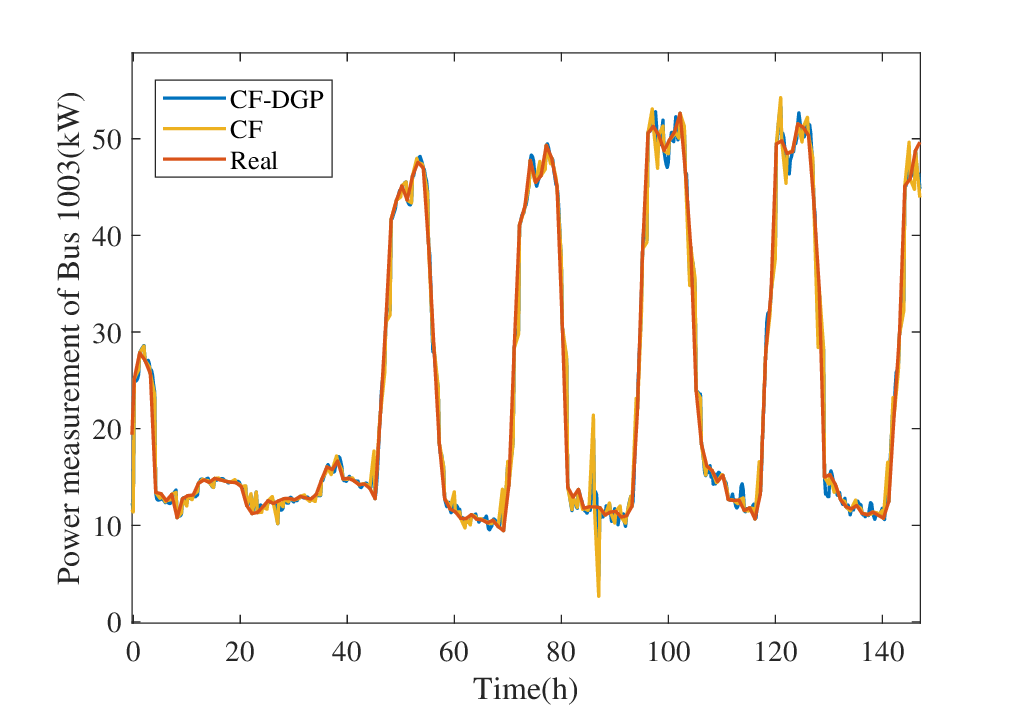}
    \vspace{-2em}
    \caption{\small The performance comparison of CF and CF-DGP}\label{fig:compa}
  \end{figure}
  
  \begin{figure}[ht]
    \centering
    \includegraphics[width=0.5\textwidth]{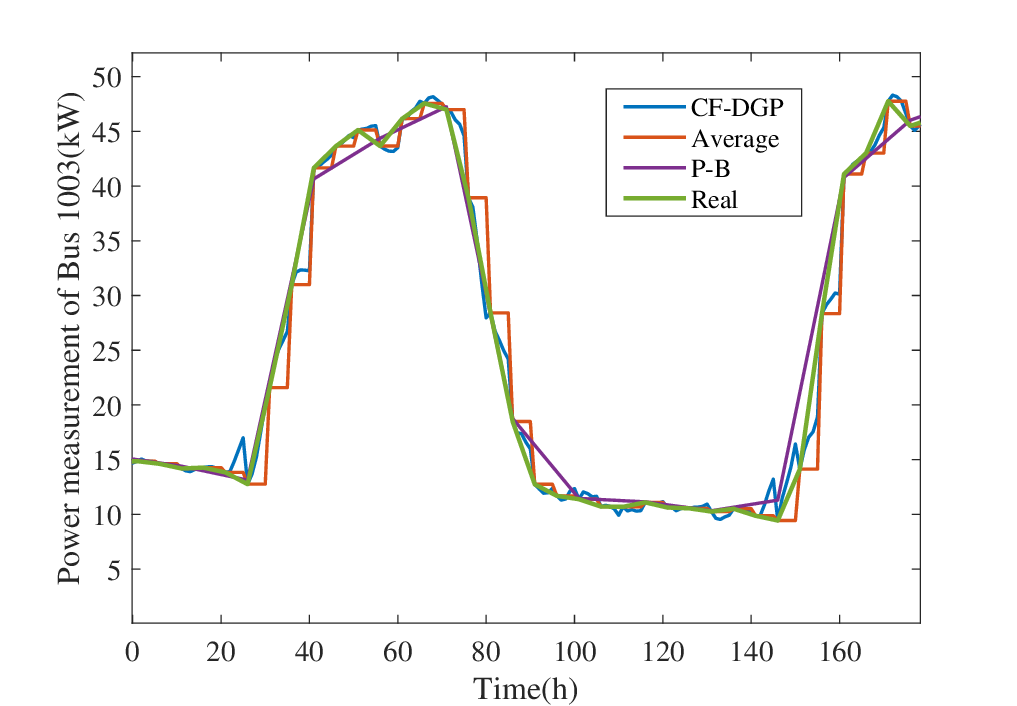}
    \vspace{-2em}
    \caption{\small The performance comparison of CF-DGP and other methods}\label{fig:compa2}
  \end{figure}
  To illustrate the effectiveness of the proposed CF-DGP method, we compared it with i)CF given in (\ref{eq:CF}), ii)the average power(the cumulative energy consumption divided by $T$ intervals) and iii)the prediction-based(P-B) method(predict the smart meter measurements and employ linear interpolation to generate pseudo measurements). Since the true system states are available in this case, we directly compare the state estimation results of different methods with the true power injection, using the following index:
  \begin{equation}
  \mathrm{RMSE}\!=\!\sqrt{\frac{\sum_{i=1}^N(\hat{p}_i\!-\!p_i)^2}{N}}
  \end{equation}
  where $\hat{p}_i$ and $p_i$ are the estimated power injection and the true injection at time $i$, respectively. $N$ is the number of sampling points. We also adopt MAPE as the evaluation metric, with a similar definition in \eqref{eq:MAPE}. The comparison of performance indices is shown in Table \ref{table:compa}.
  
  Generally, the proposed CF-DGP method shows preponderance over other methods. As is shown in Fig.\ref{fig:compa}, compared with the CF method, the CF-DGP is more robust with abnormal events. The traditional CF method will cause spikes when there are gross errors in correlated measurements. Fig.\ref{fig:compa2} shows the estimated fast-scale measurements of CF-DGP, P-B and average method. It can be concluded that the CF-DGP fits the true power injection better compared with other methods. The prediction-based method exhibits good performance regarding RMSE, while its MAPE is not satisfying. We would like to emphasize that the CF has not been applied with smart meters in the literature hitherto and using CF only is already more precise than the existing methods.
  
  \begin{table}[ht]
  \renewcommand{\arraystretch}{1.2}
  \centering
  \caption{RMSE and MAPE achieved by different methods}
  \setlength{\tabcolsep}{8mm}{
  \begin{tabular}{lll}
  \hline
  Approaches&RMSE&MAPE\\
  \hline
  Average&2.7253&5.3296\%\\
  P-B&2.0715&6.7099\%\\
  CF&2.0146&4.9073\%\\
  CF-DGP&1.7883&4.6708\%\\
  \hline
  \end{tabular}\label{table:compa}}
  \end{table}
  
  \section{Conclusion}
  
  A collaborative filtering approach to incorporate the smart meter data in the DSSE is proposed in this paper. The correlations of different nodes and the deep Gaussian process are used to predict the fast pseudo measurements and make the algorithm more robust. Numerical tests have shown that the RMSE and MAPE of the estimation result can be reduced by at most 34.38\% and 30.39\% with the proposed method. Future works will be on the cases of cyber-attacks and asynchronous sampling of measurements.

\bibliography{ifacconf}    

\begin{thebibliography}{19}
\providecommand{\natexlab}[1]{#1}
\providecommand{\url}[1]{\texttt{#1}}
\providecommand{\urlprefix}{URL }
\expandafter\ifx\csname urlstyle\endcsname\relax
  \providecommand{\doi}[1]{doi:\discretionary{}{}{}#1}\else
  \providecommand{\doi}{doi:\discretionary{}{}{}\begingroup
  \urlstyle{rm}\Url}\fi

\bibitem[{Abur and Exposito(2004)}]{abur2004power}
Abur, A. and Exposito, A.G. (2004).
\newblock \emph{Power system state estimation: theory and implementation}.
\newblock CRC press.

\bibitem[{Alimardani et~al.(2015)Alimardani, Therrien, Atanackovic, Jatskevich,
  and Vaahedi}]{alimardani2015distribution}
Alimardani, A., Therrien, F., Atanackovic, D., Jatskevich, J., and Vaahedi, E.
  (2015).
\newblock Distribution system state estimation based on nonsynchronized smart
  meters.
\newblock \emph{IEEE Transactions on Smart Grid}, 6(6), 2919--2928.

\bibitem[{Bu et~al.(2019)Bu, Yuan, Wang, Dehghanpour, and Kimber}]{bu2019time}
Bu, F., Yuan, Y., Wang, Z., Dehghanpour, K., and Kimber, A. (2019).
\newblock A time-series distribution test system based on real utility data.
\newblock In \emph{2019 North American Power Symposium (NAPS)}, 1--6. IEEE.

\bibitem[{Cao et~al.(2021)Cao, Zhao, Hu, Zhang, Liao, Chen, and
  Blaabjerg}]{cao2021robust}
Cao, D., Zhao, J., Hu, W., Zhang, Y., Liao, Q., Chen, Z., and Blaabjerg, F.
  (2021).
\newblock Robust deep gaussian process-based probabilistic electrical load
  forecasting against anomalous events.
\newblock \emph{IEEE Transactions on Industrial Informatics}, 18(2),
  1142--1153.

\bibitem[{Damianou and Lawrence(2013)}]{damianou2013deep}
Damianou, A. and Lawrence, N.D. (2013).
\newblock Deep gaussian processes.
\newblock In \emph{Artificial intelligence and statistics}, 207--215. PMLR.

\bibitem[{Do~Coutto~Filho and de~Souza(2009)}]{do2009forecasting}
Do~Coutto~Filho, M.B. and de~Souza, J.C.S. (2009).
\newblock Forecasting-aided state estimation—part i: Panorama.
\newblock \emph{IEEE Transactions on Power Systems}, 24(4), 1667--1677.

\bibitem[{G{\'o}mez-Exp{\'o}sito et~al.(2014)G{\'o}mez-Exp{\'o}sito,
  G{\'o}mez-Quiles, and D{\v{z}}afi{\'c}}]{gomez2014state}
G{\'o}mez-Exp{\'o}sito, A., G{\'o}mez-Quiles, C., and D{\v{z}}afi{\'c}, I.
  (2014).
\newblock State estimation in two time scales for smart distribution systems.
\newblock \emph{IEEE Transactions on Smart Grid}, 6(1), 421--430.

\bibitem[{Hayes et~al.(2014)Hayes, Gruber, and Prodanovic}]{hayes2014closed}
Hayes, B.P., Gruber, J.K., and Prodanovic, M. (2014).
\newblock A closed-loop state estimation tool for mv network monitoring and
  operation.
\newblock \emph{IEEE Transactions on Smart Grid}, 6(4), 2116--2125.

\bibitem[{Li et~al.(2019)Li, Sun, Wang, Zhou, and Lin}]{li2019enhanced}
Li, L.L., Sun, J., Wang, C.H., Zhou, Y.T., and Lin, K.P. (2019).
\newblock Enhanced gaussian process mixture model for short-term electric load
  forecasting.
\newblock \emph{Information Sciences}, 477, 386--398.

\bibitem[{Mestav et~al.(2019)Mestav, Luengo-Rozas, and
  Tong}]{mestav2019bayesian}
Mestav, K.R., Luengo-Rozas, J., and Tong, L. (2019).
\newblock Bayesian state estimation for unobservable distribution systems via
  deep learning.
\newblock \emph{IEEE Transactions on Power Systems}, 34(6), 4910--4920.

\bibitem[{Pareek and Nguyen(2020)}]{pareek2020gaussian}
Pareek, P. and Nguyen, H.D. (2020).
\newblock Gaussian process learning-based probabilistic optimal power flow.
\newblock \emph{IEEE Transactions on Power Systems}, 36(1), 541--544.

\bibitem[{Primadianto and Lu(2016)}]{primadianto2016review}
Primadianto, A. and Lu, C.N. (2016).
\newblock A review on distribution system state estimation.
\newblock \emph{IEEE Transactions on Power Systems}, 32(5), 3875--3883.

\bibitem[{Salimbeni and Deisenroth(2017)}]{salimbeni2017doubly}
Salimbeni, H. and Deisenroth, M. (2017).
\newblock Doubly stochastic variational inference for deep gaussian processes.
\newblock \emph{Advances in neural information processing systems}, 30.

\bibitem[{Singh et~al.(2009)Singh, Pal, and Jabr}]{singh2009statistical}
Singh, R., Pal, B.C., and Jabr, R.A. (2009).
\newblock Statistical representation of distribution system loads using
  gaussian mixture model.
\newblock \emph{IEEE Transactions on Power Systems}, 25(1), 29--37.

\bibitem[{Snelson and Ghahramani(2006)}]{snelson2006sparse}
Snelson, E. and Ghahramani, Z. (2006).
\newblock Sparse gaussian processes using pseudo-inputs.
\newblock \emph{Advances in neural information processing systems}, 18, 1257.

\bibitem[{Su and Khoshgoftaar(2009)}]{su2009survey}
Su, X. and Khoshgoftaar, T.M. (2009).
\newblock A survey of collaborative filtering techniques.
\newblock \emph{Advances in artificial intelligence}, 2009.

\bibitem[{Wu et~al.(2012)Wu, He, and Jenkins}]{wu2012robust}
Wu, J., He, Y., and Jenkins, N. (2012).
\newblock A robust state estimator for medium voltage distribution networks.
\newblock \emph{IEEE Transactions on Power Systems}, 28(2), 1008--1016.

\bibitem[{Ye et~al.(2021)Ye, Zhao, Ding, Yang, Chen, and
  Dobbins}]{ye2021global}
Ye, K., Zhao, J., Ding, F., Yang, R., Chen, X., and Dobbins, G. (2021).
\newblock Global sensitivity analysis of large distribution system with pvs
  using deep gaussian process.
\newblock \emph{IEEE Transactions on Power Systems}.

\bibitem[{Zhang et~al.(2019)Zhang, Meng, Kong, and Dong}]{8412100}
Zhang, Y., Meng, K., Kong, W., and Dong, Z.Y. (2019).
\newblock Collaborative filtering-based electricity plan recommender system.
\newblock \emph{IEEE Transactions on Industrial Informatics}, 15(3),
  1393--1404.
\newblock \doi{10.1109/TII.2018.2856842}.

\end{thebibliography}
\end{document}